**Regulating Next-Generation Implantable Brain-Computer Interfaces: Recommendations for Ethical Development and Implementation**

Renée A. Sirbu,[1] Jessica Morley,[1] Tyler Schroder,[1,2] Raghavendra Pradyumna Pothukuchi,[2] Muhammed Ugur,[2] Abhishek Bhattacharjee,[2] Luciano Floridi[1,3]

[1] Yale Digital Ethics Center, Yale University, 85 Trumbull Street, New Haven, CT 06511, US

[2] Department of Computer Science, School of Engineering and Applied Science, Yale University, 51 Prospect St., New Haven, CT 06511, US

[3] Department of Legal Studies, University of Bologna, Via Zamboni 22, Bologna, 40100, IT

Email for correspondence: renee.sirbu@yale.edu

**Abstract**

Brain-computer interfaces (BCIs) offer significant therapeutic opportunities for a variety of neurophysiological and neuropsychiatric disorders and may perhaps one day lead to augmenting the cognition and decision-making of the healthy brain. However, existing regulatory frameworks designed for implantable medical devices (IMDs) are inadequate to address the unique ethical, legal, and social risks associated with next-generation networked brain-computer interfaces (BCIs). In this article, we make nine recommendations to support developers in the design of BCIs and nine recommendations to support policymakers in the application of BCIs, drawing insights from the regulatory history of IMDs and principles from AI ethics. We begin by outlining the historical development of IMDs and the regulatory milestones that have shaped their oversight. Next, we summarize similarities between IMDs and emerging implantable BCIs, identifying existing provisions for their regulation. We then use two case studies of emerging cutting-edge BCIs—the HALO (Hardware Architecture for LOw-power BCIs) and SCALO (SCalable Architecture for LOw-power BCIs) computer systems—to highlight distinctive features in the design and application of next-generation BCIs arising from contemporary chip architectures, which necessitate reevaluating regulatory approaches. We identify critical ethical considerations for these BCIs, including unique conceptions of autonomy, identity, and mental privacy. Based on these insights, we suggest potential avenues for the ethical regulation of BCIs, emphasizing the importance of interdisciplinary collaboration and proactive mitigation of potential harms. The goal is to support the responsible design and application of new BCIs, ensuring their safe and ethical integration into medical practice.

## 1. Introduction

Improvements in transistor technology led to the widespread adoption of implantable medical devices (IMDs) throughout the twentieth century. Following these developments, the modern era of IMDs can support more sophisticated electronic and computing capabilities than their predecessors. This trend has revolutionized the treatment of countless medical conditions, from cardiac arrhythmias to diabetes. However, IMDs also introduced new risks, and, in response, the US Food and Drug Administration (FDA) established a comprehensive framework for their regulation in 1976, known as the Medical Device Amendments (MDA). The MDA sought to update the Food, Drug, and Cosmetic (FD&C) Act of 1938 to include authority for the premarket approval of medical devices.[1] Since then, most IMDs have been classified under the FD&C Act as Class III, or "high-risk," devices.[1] Thus, IMDs have been subjected to the strictest possible regulatory scrutiny for nearly five decades, ensuring that patients can benefit from their therapeutic potential while being protected from harm. The advent of new IMDs, such as next-generation implantable brain-computer interfaces (BCIs), poses a risk of exposing inadequacies in the existing regulatory framework.

The first BCI was successfully developed in 1963, marking the beginning of a new frontier in medical device technology.[2] The IEEE defines any BCI as "a system that establishes a direct communication channel between the human or animal brain and a computer or external device," which "[records] or [stimulates] activity of the central or peripheral nervous system (CNS/PNS) to replace, restore, supplement, or improve natural output/input."[4] BCIs can be non-invasive (wearable) or invasive (implantable). Implantable BCIs are often preferable to wearable BCIs, especially in medical use cases, because they can stimulate and record "large numbers of neurons with high signal fidelity, spatial resolution, and in real-time," compared to wearables, which produce noisier, lower-resolution signals.[4] High spatial and temporal signal fidelity is vital in treating neurological and neuropsychiatric disorders. BCIs have been used as assistive devices for patients with severe motion-limiting disabilities, seizure disorders, and treatment-resistant mental illnesses, among many other neurophysiological and mental disorders.[5] These devices offer significant therapeutic opportunities, but their use introduces new ethical, legal, and social risks that existing IMD regulations are ill-equipped to address.

BCIs differ from early-generation IMDs and conventional computing systems because of the vulnerability of their implanted environment, which creates more acute power, energy, and thermal constraints than those typically faced by hardware designers building server and mobile systems. For example, BCI hardware designers must balance electronic and engineering principles with the chip's biological compatibility within the brain tissue. The power budget of a BCI is constrained, as brain tissue can only withstand heating of up to one degree Celsius, diffused by approximately tens of milliwatts of power (depending on the target region), before cellular damage occurs.[4] Thus, hardware must be compact, durable, and safe for implantation.

---

[1] These include only life-sustaining or life-saving implantable devices; according to the FDA database, 404 Class II IMDs currently exist.

Furthermore, the intimate connection between these devices and the human brain raises unprecedented ethical questions about preserving patient autonomy, identity, and mental privacy. As BCIs continue to advance, incorporating artificial intelligence and machine learning (AI/ML),[6,[2]] and networked architectures,[4,7,8] the gap between existing regulatory frameworks and the capabilities of these devices widens. Thus, in this article, we aim to help close this gap by providing six recommendations to support developers in the design of BCIs and nine recommendations to support policymakers in the ethical application of BCIs. We argue that existing regulatory frameworks for IMDs can be amended[3] according to these recommendations to address the unique ethical, legal, and social risks posed by BCIs. The rest of the article is structured as follows.

In Section 2, we introduce our methodology. In Section 3, we compare the similarities and identify the differences between traditional IMDs and next-generation BCIs to highlight gaps in current regulatory frameworks. Then, we select an ethical framework to help structure our recommendations for addressing these gaps later. In Section 4, we examine the landscape of existing BCI technologies, with an empirical focus on the HALO (Hardware Architecture for Low-power BCIs)[7] computer system for BCIs, and its successor, the SCALO (SCalable Architecture for LOw-power BCIs) computer system.[8] In Section 5, we consider how the SCALO model may serve as evidence of the widening gap between existing IMD regulations and the increasingly sophisticated capabilities of BCIs. We apply our methodology to formulate nine recommendations for BCI developers, targeting specific design elements that SCALO already employs and for which regulations must be amended accordingly. In Section 6, we identify the salient ethical, legal, and social risks associated with next-generation BCIs, such as SCALO, and provide nine recommendations to policymakers that address these risks to protect stakeholder interests. Finally, in Section 7, we conclude and discuss the complexities of regulating increasingly advanced human-machine interfaces.

---

[2] Existing literature (including existing frameworks advanced by the FDA in the oversight of new IMDs and software-as-a-medical device) couples AI and ML as "AI/ML" in accepted definitions. We thus keep this semantic coupling for consistency with current regulatory structures.
[3] We use "amended" throughout to refer to both *expanding* existing frameworks, in some cases, and *updating* existing frameworks, in others.

## 2. Methodology

Our analysis compares the similarities and identifies differences between traditional IMDs and BCIs to identify gaps in current regulatory frameworks. We structure our analysis around the principles outlined in Floridi and Cowls's *A Unified Framework of Five Principles for AI in Society* (2019), henceforth referred to as the Unified Framework: beneficence, non-maleficence, autonomy, justice, and explicability.[9] This framework standardizes ethical guidance advanced by international governing bodies in the AI regulatory landscape.[9,10] We use this framework for three reasons. First, it accounts for the core principles of bioethics (beneficence, non-maleficence, autonomy, and justice)[11,12] relevant to the ethical regulation of medical devices as emphasized by the American Medical Association (AMA).[13] Second, it extends these tenets to encompass *explicability*, a unique consideration required for the ethical development of new AI technologies.[14] [4] Third, the Unified Framework is based on an analysis of six sets of principle documents,[15] all of which were summarized to fall within these five umbrella principles. This approach enables us to address the ethical risks of BCIs systematically.

Despite increased public speculation, many applications of BCIs require technology that is still unavailable, untested, or unsafe. The Royal Society *iHuman perspective: Neural interfaces* report (2019) identifies a range of current, near-term, and longer-term BCI applications spanning clinical treatment of neurological disorders, entertainment, leisure, and even augmentation of the healthy brain. Some of these applications include the mental navigation of computers and other mouse-and-keyboard devices, direct brain-to-brain communication, continuous monitoring of neural activity, and the enhancement of human intellectual faculties (e.g., memory, concentration, and learning).[6] However, among these use cases, implantable BCIs are most likely to be deployed for disease treatment.[5] We focus on these latter use cases via two BCI architectures developed at Yale University: HALO[7] and its successor, SCALO.[8] HALO and SCALO inform our discussion of next-generation design and application considerations. In particular, we focus on SCALO as an example of a networked, next-generation BCI that challenges existing regulatory frameworks to highlight design considerations and applications that regulators should prioritize.

---

[4] See *Appendix I* for a comparison of "A Unified Framework of Five Principles for AI in Society" and the principles of bioethics relevant to the ethical regulation of medical devices, as identified by the AMA.

[5] Since the conception of this article, media attention has surfaced surrounding Neuralink's first implanted human BCI and the patient's ability to perform mouse-and-keyboard navigation using brain control. We stress that Neuralink's trial has not been recorded in ClinicalTrials.gov, or in any public repository required for IRB approval, so we have not included its analysis in this article.

### 3. Similarities and differences between BCIs and other Class III IMDs

Before identifying the unique ethical, legal, and social risks of next-generation implantable BCIs, it is essential to understand their functional similarities with existing Class III IMDs.[6] Many functional similarities between traditional IMDs and BCIs can be explained by their nature as artificial, technological devices, implanted into a biological organism. These devices: (i) differ from transplanted organs; (ii) require surgical intervention for implantation; (iii) can be used as bridging or destination therapies;[16] (iv) must be removed in the case of manufacturer bankruptcy or company dissolution;[17] (v) introduce new data privacy questions; (vi) introduce new approval pathway questions if they employ AI/ML software;[18] and (vii) are primarily intended to serve as therapeutic alternatives in high-risk or otherwise treatment-resistant cases. Existing regulations have thus been advanced to address these similarities. Table 1 provides a non-exhaustive overview of the functional similarities between BCIs and other Class III IMDs, as well as their existing oversight, in alignment with the principles of the Unified Framework.

Table 1: Existing regulations for Class III IMDs.

| Principle | All Class III IMDs (including BCIs) | Existing regulation(s) and guidance |
|---|---|---|
| Beneficence | Provide therapeutic benefit/disease management.[19] | FDA Class III Premarket Approval (PMA) under Title 21 CFR Part 814, Premarket Approval of Medical Devices.[20] |
| | Protect the continuity of personal identity.[19] | Fundamental human rights as defined by the Human Rights Council of the United Nations. |
| | Assure longevity and/or rechargeability of battery-operated IMDs and prevent the creation of legacy devices. | ANSI/AAMI ES 60601-1;[21]<br>Next Steps Toward Managing Legacy Medical Device Cybersecurity Risks;[22]<br>MDCG 2021-25.[23] |
| Non-maleficence | Limit access to data<br>Deidentify data and protect patient privacy. | GDPR (EU);[24]<br>HIPAA (US).[25] |

[6] The FDA defines Class III medical devices as "high risk." These devices are used for life sustaining or saving purposes and therefore require stricter regulatory controls. Examples include pacemakers or insulin pumps.

| | Protect against all four CVSS attack vectors. | Section 3305: "Ensuring Cybersecurity of Medical Devices" of the FD&C Act – Section 524B;[26]<br>CVSS v4.0 Specification Document. |
|---|---|---|
| Autonomy | Seek informed consent at implantation *and* subsequent intervention. | *Reasonable physician* standard in the informed consent doctrine of liability law;<br>FDA Informed Consent and IRB: 21 CFR 50,[27] and 56.[28] |
| | Promote autonomy in end-of-life decisions.[19] | End-of-life planning that assigns a surrogate decision maker (SDM). |
| Justice | Provide equal access to technologies;<br>Protect vulnerable groups and mitigate cultural effects.[19] | Belmont Report (1979);[12]<br>Federal Policy for the Protection of Human Subjects (Common Rule)[29] |
| Explicability | Include appropriate device labeling. | 21 CFR Part 801;[30] 820.[31] |
| | Limit direct-to-consumer marketing. | FD&C Act Section 520(e) – (21 U.S.C. 360j(e);[32]<br>Presenting Risk Information in Prescription Drug and Medical Device Promotion (Draft Guidance).[33] |

As illustrated in Table 1, existing regulations provide a foundation for the governance of BCIs, according to their functional similarities to other Class III IMDs. However, the technological sophistication of newer BCIs introduces additional considerations that require further analysis and potential regulatory updates, as discussed in the next section.

## 4. HALO, SCALO, and next-generation BCIs

The HALO chip, designed at Yale University, was built in response to concerns regarding the ability of modern BCIs to process higher-order magnitude data rates under low power constraints and with high programming flexibility.[7] HALO is an accelerator-rich processor that operates at low power (up to 15 mW) and processing rates of 46 Mbps, while permitting algorithmic flexibility and the potential for onboard personalization using programmable inter-accelerator dataflow.[7,8] HALO accomplishes this by splitting tasks into discrete functional hardware and software units that require less power than traditional ASIC designs. Despite its innovative value in balancing low power expenditure, high throughput data rate, and flexible design, HALO is designed to interface with a single brain region (although it is sufficiently flexible to accommodate that region anywhere on the cortical surface). Thus, any therapeutic benefit for specific conditions that require interfacing at multiple brain sites or the migration of treatment sites depending on neural response is limited. The SCALO (SCalable Architecture for LOw-power BCIs) chip was developed in response to these limitations, serving as the "first BCI architecture for multi-site brain interfacing in real-time."[8]

SCALO is a wirelessly distributed system of networked implants, each containing a HALO processor adapted with the necessary storage and compute architecture to support three classes of applications in a distributed manner for the first time: internal closed-loop, external closed-loop, and interactive human-in-the-loop.[8] Internal closed-loop applications modulate brain activity without external communication, responding autonomously with electrical stimulation to targeted brain sites and restoring homeostasis whenever changes are detected.[8] Some examples include the reversal of seizure onset and off-balancing tremor in Parkinson's patients.[8] External closed-loop applications modulate brain activity by communicating with systems separate from the BCI. Examples include brain-controlled devices external to the user (e.g., computer navigation) and neural prostheses (e.g., BCI-mediated prosthetic limbs).[8] Finally, interactive human-in-the-loop applications permit clinicians to adjust stimulation parameters and prompt the BCI for data, which are helpful for seizure detection, personalization of care through algorithmic tailoring, and debugging or patching an implant, for example.[8]

## 5. Nine recommendations for developers in the ethical design of BCIs

Neural activity from HALO and SCALO is processed in real-time, generating sensory, mechanical, or electrical feedback to the user,[3] bypassing traditional efferent pathways.[5,[7]] In the following subsections, we identify five potential targets for a future governance agenda concerning the responsible design of BCIs. Developers consider many trade-offs and constraints in the design of BCIs; [[8]] we advance six recommendations to address some of these trade-offs. These recommendations are meant for designers, engineers, and other back-end stakeholders involved in designing and developing next-generation implantable neurotechnologies.

---

[7] We use the traditional physiological definition of "efferent" as referring to the pathway travelled by a neuronal impulse, from nerve center (e.g., the brain) to peripheral site (e.g., the fingertips).

[8] See *Appendix II* for a table summarizing what we consider to be some of the most salient current constraints on BCI design.

### 5.1. Non-invasive updating and hardware flexibility

All invasive BCIs must support onboard hardware flexibility to perform wireless updates.[4] Otherwise, manual patching, debugging, and updating, which require surgical intervention, introduce undue risks.[34] Thus, chip hardware must possess sufficient memory and compute to withstand frequent, potentially large-scale updates without requiring manual intervention or replacement. This also requires hardware to be wirelessly compatible, engaged in different locales, and actionable in different deployment scenarios. Additionally, the implant must handle an update failure gracefully; it cannot "brick" and require surgery to reset.[34,[9]]

**Recommendations**

(1) Ensure onboard flexibility and sustainability throughout the device's lifespan (no less than 12-15 years, in line with existing IMD lifecycle standards).[35]

(2) Support human-in-the-loop applications for patchwork.

(3) Ensure that wireless capabilities are widely accessible and compatible with common devices and technologies.

**Rationale**

Reducing the amount of surgical intervention necessary for the functional maintenance of next-generation BCIs would improve patient quality of life. [36] Wireless compatibility is essential to prevent surgical intervention and ensure seamless updates. The NeuroVista system serves as an example of the potential harm that can arise from proprietary wireless access or systems lacking wireless compatibility altogether.[37] NeuroVista required a custom handheld device for continuous monitoring and updates; the company supplying the device went bankrupt, limiting the long-term functionality of the implant and any possibility for non-invasive updates. Scenarios like this can be avoided by implementing the three recommendations described above.[38]

Non-invasive updating is a necessary function of next-generation implantable BCIs, because onboard software must be patient-specific and able to handle neural variability over time. Furthermore, since newer BCIs increasingly rely on AI/ML models,[6] and adaptive learning is an inherent component of many of these models, these devices could require frequent updates to keep pace with each software iteration. Hands-off improvements and non-invasive updates can be achieved by increasing device battery life and improving biological viability through materials research (to delay hardware replacement), or enhancing interoperability between the device and control software for the end-user.[39] The SCALO chip provides a model architecture to support the relevant applications (e.g., human-in-the-loop) required for off-board patchwork and debugging.[8] Deploying more flexible chip hardware in next-generation BCIs satisfies the principles of beneficence and non-maleficence by ensuring long-term device functionality while minimizing the need for invasive interventions.

---

[9] A non-responsive, unusable state for hardware; incapable of proceeding without intervention (and not easily reversible).

**Technical Considerations**

Supporting hardware flexibility comes at the expense of safer designs. More flexible hardware is known to increase power consumption, which may exceed safe implantation limits. This flexibility will naturally be more complex and, thus, more challenging to design and test. Thus, the method by which to update the device noninvasively must be handled with care. Lower-overhead techniques for updating on-device software and parameters are preferred, which may require tightening the hardware-software interface.

Designers must also navigate a wide range of radio technologies and communication options for a BCI, which will impact how, when, and in what conditions a device may be wirelessly engaged. The Federal Communications Commission (FCC) and the European Telecommunications Standards Institute (ETSI) define standards and operating conditions for various radio protocols, which limit the range of communication, the environment in which the device can be used, and the amount of data transferred wirelessly within a given period, taking into account power constraints. These standards will define how and when non-invasive updating will be performed. Ensuring that the device is compatible with standard protocols, such as Bluetooth, may be sufficient to handle updates. However, if higher data rates are required, the wireless subsystem may become more customized and less compatible, limiting accessibility and interoperability.

## 5.2. Algorithmic personalization and programmability

BCIs require greater software flexibility for many reasons. Neurological and psychiatric disorders vary from person to person; the brain's day-to-day plasticity leads to variability;[40] sensor degradation reduces signal quality due to the immune system's response to implanted electrodes;[41,42] and clinical best practices are constantly evolving.[43] Thus, next-generation BCI software must be adaptable enough to respond to these biological and physical changes during the implant's operational life, with the ultimate goal of covering a patient's lifespan. This will ensure that the patient receives adequate treatment as their condition(s) progress over time.

**Recommendations**

(4) Permit multi-site neuromodulation within power budgets, enabling individual brain sites to receive closed-loop stimulation in varying amounts, tailored to the patient's needs.

(5) Enable flexible neural recording frontends to filter out unwanted features and digitize only what is necessary for treatment.

**Rationale**

Multi-site neuromodulation enhances software flexibility by enabling stimuli to adapt in response to individual or demographic differences.[8,44] Specifically, this function supports variability in the level of stimulation deployed to independent brain regions, even among patients with the same diagnosis. These impulses can also vary in strength depending on the intensity of the symptoms. Flexible recording frontends will allow for greater personalization by removing noise, artifacts, and unwanted frequency ranges. They can be configured to handle just enough precision for the intended treatment. Thus, multi-site neuromodulation and programmable frontends, coupled with onboard flexibility and human-in-the-loop patchwork as previously defined, would permit personalization over time without requiring invasive corrective procedures. Furthermore, greater software flexibility and programmability would allow AI/ML-enhanced BCIs to adapt to iterative learning processes.[6] Networked BCIs, such as SCALO, support a higher level of software flexibility because signals elicited by independent implanted regions necessarily trigger variable feedback responses.[8] Design architectures that support multi-site neuromodulation and programmable recording frontends align with the principles of beneficence and justice by accounting for individual and demographic differences to promote personalized treatment options.

**Technical Considerations**

Enabling personalization will require sufficient software flexibility and programmability, allowing decoders and stimulation parameters to be adapted on demand. These parameters should furthermore be adjustable to the needs of different end users. Clinicians, neuroscientists, caretakers, and other stakeholders may not be well-versed in low-level embedded programming that may be required to configure and update these devices over time. Thus, designers should expose higher-level interfaces and programming paradigms to authorized users, enabling them to use the offered flexibility effectively. Otherwise, managing individual differences in treatment, evolution of algorithms, neural plasticity, and degradation over time will prove to be a burden and hinder timely treatment.

### 5.3. Closed-loop mechanics

In addition to response variability, response *constancy* is a target for improving BCI design. The open-loop mechanics of early-generation BCIs involved deploying constant stimuli to target regions, independent of feedback signals from their implant sites.[45] However, several concerns have been raised regarding these open-loop systems, including their overall treatment value.[45,46] BCIs that operate in a closed loop with minimal reliance on the network will improve responsiveness and provide timely treatment.

**Recommendation**

(6) Stress closed-loop design in networked BCIs, wherein each unit (i.e., chip) can operate within its feedback loop, and the system at large can operate within a monolithic closed-loop and continue to function independently of a central control chip.

**Rationale**

Closed-loop systems are more aligned with modern therapeutic goals, supporting feedback-dependent stimulation through bidirectional communication between the brain site and the device, as well as multi-site neuromodulation.[47] These systems are increasingly preferred to their open-loop predecessors for several reasons, including environmental impact, treatment personalization, and overall treatment potential.[46] For example, feedback-dependent (i.e., non-constant) stimulation helps preserve battery life, meaning devices can last longer on a single charge and produce less waste.[45]

Prolonged battery life also results in less manual intervention, whether through an extended time between hardware replacement surgeries or a reduced active burden on the BCI recipient themselves (in cases where inductive charging of the BCI requires the patient to recharge the implant via external electrodes).[48] Furthermore, in closed-loop, networked BCI models like SCALO, individual units can operate independently of a central control chip in the event of communication failure or interference, meaning that even if one unit is affected, the system as a whole can continue to operate.[8,47] Thus, this improvement alone may improve patient quality of life while supporting environmental sustainability initiatives.

Perhaps the most significant value of "closing the loop" arises from treating complex neuropsychiatric conditions via deep brain stimulation (DBS) and similar therapies.[46] Emerging research suggests that disorders like Parkinson's may respond better to targeted neuromodulation (stimulation only delivered in response to neural signals indicating a tremor).[46] Thus, closed-loop systems align with the principles of non-maleficence—by reducing the need for surgical interventions and battery replacements while improving overall treatment outcomes and reducing environmental

waste—and autonomy—by reducing the frequency of medical visits such that patients with closed-loop BCI systems would be able to continue daily activities without regular interruption.

**Technical Considerations**

Devices that have reduced reliance on the network for sending and receiving data during operation will consume less energy and become more responsive. The network may have its own latency and connectivity concerns, which may prove to be less reliable compared to fully closed-loop systems. Furthermore, system-level independence (and concurrency) is only one aspect of BCI design. Concurrency among components within a single device is also necessary for adequately responding to brain events. For example, sensing, stimulation, and wireless communication for both transmitting and receiving (i.e., complete duplex systems) should happen concurrently, if necessary.

### 5.4. Data encryption and storage

The HALO study identifies seven contemporary BCIs that do not encrypt their data storage or transmission processes, directly opposing the HIPAA Privacy Rule.[7,25] Historically, there have been two reasons for this lack of integration. First, the ultra-low power budget available to early-generation BCIs did not support encryption without limiting the implant's therapeutic capabilities.[45] For older devices, second, off-device data transmission, storage, or translation was unnecessary.[49] In contrast, newer BCIs are moving towards closed-loop architectures (see *§3.3*) and increasingly employing off-device AI/ML processing to (a) adapt to neural feedback signals and (b) translate brain data for use by clinicians.[45,47,48]

**Recommendation**

(7) Introduce in-transit encryption measures wherever power constraints still permit the full functionality of the device.

(8) Integrate on-device storage for flexible data management and privacy.

**Rationale**

For early-generation BCIs, developers considered encryption "at rest," meaning encrypting data stored on the chip in short- or long-term storage.[4] However, due to the nature of the implanted environment, a physical cyberattack is less likely for a BCI than a traditional computing system simply because it is far more challenging to access a brain than a hard drive.[10] On-device encryption is thus not an appropriate solution to next-generation BCI security concerns, since it does not protect *off-device* data transmission.[50] Instead, *in-transit* encryption is quickly becoming imperative for the ethical design of newer BCIs. Low-power chips like SCALO can support in-transit encryption between nodes and remote entities, meaning that data are encrypted as they are transmitted off-device and onto the network.[8] And, since SCALO's networked structure offers some respite from tight power constraints, it encourages better measures of encryption that adhere to existing privacy regulations for protecting personal health information (PHI), following HIPAA protocol. Thus, introducing appropriate in-transit encryption into new BCI models addresses the principle of non-maleficence by protecting patient data from unauthorized access or breaches. Furthermore, on-device storage, typically in the form of non-volatile memory, can record events and track data over extended periods before sending the information off-device. This can improve data privacy by limiting data sent off-device and keeping most data local.

---

[10] Some BCIs (and IMDs) contain physical SD ports, making physical attacks *possible*, though still significantly less likely than traditional computing systems.

**Technical Considerations**

Encryption on-device can be computationally intensive and consume excess power. Custom circuits for encryption will be necessary, and the choice of encryption scheme will be crucial for both regulatory compliance and power, performance, and chip area considerations. Methods for reducing overhead by co-designing the communication layers with the encryption scheme are preferable.

In terms of storage, there are many possible non-volatile memory technologies to choose from, each with its device lifetime.[51] These technologies have a fixed number of times they can be written to before becoming unusable. Designers must be cautious of overwriting data to prolong lifespan and avoid replacement.[52]

### 5.5. Network connectivity and access

The constrained functionality of traditional IMDs and their ability to run "unsupervised" for prolonged periods meant that network disconnection had a negligible effect on their operation.[49] Wired cardiac pacemakers, for example, respond to physiological signals transmitted by cardiac tissue through flexible, insulated leads and into implanted electrodes, and wireless cardiac pacemakers deliver constant, modulating pulses directly into the right ventricle.[53] As described in *§3.2*, many of these IMDs (including early-generation BCIs) did not require the storage or transmission of physiological data to function successfully. However, unlike cardiac pacemakers and early-generation BCIs, the successful function of newer BCIs relies on their ability to review, store, and respond to brain data.[54] As such, Internet of Things (IoT) connectivity is integral to the operation of these devices.

**Recommendation**

(9) Explore authorization schemes that permit network access only to authorized parties.

**Rationale**

Since next-generation BCIs, such as SCALO, cannot be simply disconnected from the IoT, maintaining trust between stakeholders is imperative for the ethical adoption of these devices. Developers should maintain high standards for access, either through in-transit data encryption measures or robust cybersecurity protocols. [11] Furthermore, clinicians should have limited access to patient data *as these data are required for treatment* (e.g., in monitoring outcomes, but not around-the-clock surveillance). Patients should feel empowered to provide or withhold consent to their data along the care continuum (not just at the point of implantation) and be actively aware of which PHI data are being collected. Limiting authorized access supports the principles of autonomy, justice, and explicability by ensuring that patients understand and control their data, while still allowing necessary access for therapeutic purposes.

**Technical Considerations**

Authorization will require subjecting permitted receiving and transmitting devices to a communication protocol. The choice of protocol is currently not standardized and will have a significant impact on the power and performance of any authorization scheme or mechanism. Only devices that implement the defined protocol will be candidates for authorization. Other layers of the network stack will be required to implement the specifics of the authorization scheme. As previously described, on-device storage can help mitigate frequent network disconnections, allowing data to be stored without needing to send it over the network and then accessed once connectivity is resumed. The various storage options will impact this functionality and must be taken into consideration.

Table 2 summarizes the novel design elements identified in this section and the recommendations advanced in their consideration.

[11] Given the sensitive nature of both the implantation site and the data collected, many design considerations require enhanced cybersecurity protocols. Some of these protocols are introduced in subsequent analysis but will be further developed in a comprehensive companion article.

Table 2: Summary – Design

| Considerations for next-generation BCI *design* | Recommendation(s) | Principle(s) addressed |
|---|---|---|
| Non-invasive patchwork, debugging, and software updating | 1. Ensure onboard flexibility and sustainability over the device lifespan. | Beneficence; Non-maleficence. |
| | 2. Support human-in-the-loop applications for patchwork. | Beneficence; Non-maleficence. |
| | 3. Ensure that wireless capabilities are widely accessible and compatible with common devices and technologies. | Autonomy; Beneficence. |
| Algorithmic personalization and software flexibility | 4. Permit multi-site neuromodulation within power budgets, allowing individual brain sites to receive closed-loop stimulation in varying amounts, tailored to the patient's needs. | Beneficence; Justice. |
| | 5. Enable flexible neural recording frontends to filter out unwanted features and digitize only what is necessary for treatment. | Beneficence |
| Closed-loop mechanics | 6. Stress closed-loop design in networked BCIs, wherein each unit (i.e., chip) can operate within its feedback loop, and the system at large can operate within a monolithic closed-loop and continue to function independently of a central control chip. | Non-maleficence; Autonomy. |
| Data encryption and storage | 7. Introduce in-transit encryption measures wherever power constraints still permit the full functionality of the device. | Non-maleficence. |
| | 8. Integrate on-device storage for flexible data management and privacy | Beneficence |
| Network connectivity and access | 9. Explore authorization schemes that permit network access only to authorized parties. | Autonomy; Justice; Explicability. |

## 6. Nine recommendations for policymakers in the ethical application of BCIs

BCIs can stimulate movement in paralyzed limbs, restore vision, and reverse seizure onset, among other potential therapeutic applications.[6] However, their use may also extend beyond treatment, introducing challenging new questions regarding human autonomy in various contexts, including agency, entertainment, and even government action.[6,55] For instance, introducing a foreign computing system into the brain—often considered the locus of identity or the "final frontier" of privacy—raises questions surrounding autonomous decision-making, namely, whether conscious choice *can* exist in the face of BCI integration.[5,56,57] The implications of BCI use on gaming and entertainment are also significant, altering the nature of interaction between human game players and between humans and machines, and introducing new concerns surrounding skill enhancement (e.g., improvements in vision and reaction time).[6] Furthermore, a citizen's right to privacy and freedom of thought may be jeopardized depending on the government's access to personal devices, especially in non-democratic nations.[58] For example, BCIs can facilitate direct human-human neural communication (i.e., unmediated by external communication devices) and paired human-machine decision-making (i.e., actioning a command via neural signals)—points of significant interest for both international governing bodies and independent states.[59]

In the following subsections, we identify seven target areas for a future governance agenda related to the ethical application of next-generation BCIs, such as SCALO. We present nine recommendations across seven target areas to policymakers, regulators, international governing bodies, and other key stakeholders involved in the deployment of next-generation implantable neurotechnologies.

### 6.1. Restoration of function to "normal" threshold

The physiological bounds of "normal organ function" are usually empirically determined and measured by available clinical means, such as laboratory tests. For organs like the liver, measures that fall outside of specified parameters could indicate significant health problems, like cirrhosis or hepatitis. However, definitions of "normal thresholds" for brain function are not always clearly quantitatively defined.

**Recommendations**

(1) Maintain quantitative threshold standards for neurophysiological illnesses with sufficient reference data (e.g., Parkinson's disorder) and permit neurostimulation up until "normal" thresholds.

(2) Offer deep brain stimulation (DBS) by BCI-mediated action for treatment-resistant neuropsychiatric disorders, though not as a first-pass therapeutic intervention.
(3) Couple BCI treatment with cognitive behavioural therapy (CBT), dialectical behaviour therapy (DBT), eye movement desensitization and reprocessing (EMDR), or other forms of clinically accepted therapy under a licensed practitioner.

**Rationale**

In some cases, particularly those relating to mental states or regarding a mental health diagnosis, individuals may define "normal thresholds" differently based on their life narratives, values, and personal expectations, which are not easily quantified by laboratory testing. Defining "normal thresholds" in these cases may be inherently discriminatory or otherwise weaponized against vulnerable populations. However, in other cases, empirically defined parameters for specific neurological disorders are accepted among the clinical community and may be used as a reference.[60]

Balancing the therapeutic needs of individual patients with equitable treatment options for promoting population health under existing high healthcare expenditure means ensuring that (1) specific demographics are not treated (or enhanced) at the expense of others, (2) other treatment measures are exhausted before high-intensity interventions are considered, and (3) patients are supported throughout their care journey in personalized ways that reflect their goals, values, and preferences. Balancing empirical evidence with patient narrative in restoring "normal" brain function addresses the principles of beneficence, non-maleficence, and justice by providing therapeutic benefits while avoiding potential discrimination or overtreatment.

### 6.2. Enhancement of function past "normal threshold"

As previously discussed, neurological "normal thresholds" create a slippery, potentially discriminatory, slope. Despite the capacity for enhancement offered by next-generation BCIs, no regulation currently defines acceptable bounds for enhancement.

**Recommendation**

(4) Restrict BCI-mediated enhancement to special permissions.

**Rationale**

Although BCIs are not currently being used to enhance human capacity, some social groups (e.g., transhumanists, effective altruists, and tech accelerationists) have been pushing to explore this potential.[6] Existing technology may support some of these features—for instance, self-prescription (e.g., "dopamine hacking") is possible in the contemporary technological climate.[61] In this way, BCIs may provide "off-label" benefits to wealthy or dominant groups with greater access, thereby widening existing demographic disparities and further marginalizing those already marginalized populations.

Restricting enhancement mirrors the guidelines currently imposed upon pharmaceuticals with high abuse potential (i.e., controlled substances). For example, amphetamines prescribed for individuals with attention deficit hyperactivity disorder (ADHD) are also frequently taken as performance-enhancing drugs.[62] Similarly, dopamine-enhancing neuromodulation may have therapeutic benefits for treatment-resistant depression and Parkinson's, yet it may also be used for enhancement or entertainment.[63] Thus, our recommendation addresses the principles of justice and non-maleficence by preventing the misuse of BCIs for non-therapeutic purposes and centring equitable access to these technologies.

### 6.3. Decoding processes

Decoding quantitative neural activity into qualitative mental states differs from decoding cardiac activity into an electric stimulus or circulating insulin levels into a pump response. On the one hand, neural activity (e.g., epileptiform abnormalities in delta and gamma brainwaves) is measurable and can be translated into stimulus (e.g., seizure detection or reversal).[64] On the other hand, the decoding of quantitative data (i.e., brain waves) into *qualitative* characteristics (e.g., "happy," "sad," etc.) or intentions is vastly more complex.[65]

**Recommendations**

(5) Advance stakeholder-specific device labelling requirements (e.g., differing levels of expertise for manufacturer, clinician, and end-user) that center decoding functions. For example:

  a. Manufacturer labels should be fine-grained, technical, and specific, such that appropriate federal oversight can be imposed, development can be recreated, and liability (in case of malfunction) can be assigned.
  b. Clinician labels should address the potential harms of deploying BCI devices, including those posed by data collection and transmission via decoding. Clinicians should be made aware of the types of data they are analyzing, as well as the implications and limitations of these data, before issuing a treatment response.
  c. Patient labels should explain the data transmission process and how to monitor their neural activity (potentially through a user-friendly interface). Patients should be informed about any sensitive data being shared with their care team, including how they are stored, collected, and used.

**Rationale**

Since contemporary BCIs increasingly employ AI/ML software to support the decoding process, and since mental states differ between individuals and "normal thresholds" vary, decoding programs must be context-dependent and personalized. Another layer of complexity arises from translating *quantitative* data about mental states (e.g., decreased alpha waves correlating with anxiety)[66] to a *qualitative* clinical understanding, which may affect clinician response. Dedicated device labelling is envisioned as a supplement to informed consent processes, educating each stakeholder in the responsible use of these devices and empowering end-users in their care.

Labelling requirements differ among stakeholders based on the level of expertise with which they enter a clinical encounter. BCI device labelling practices should reflect these differences to promote functional understanding. This approach addresses the principles of explicability and autonomy by ensuring that all stakeholders understand how BCIs decode and interpret neural activity, thereby enabling informed decision-making and effective use.

### 6.4. Mental privacy

Mental privacy presents a unique challenge that extends beyond traditional notions of data privacy.[67] Brain data require special consideration in privacy protection frameworks, since they are inherently more intimate than other personal health information due to their potential to reveal an individual's thoughts, emotions, and cognitive processes.[5]

**Recommendation**

(6) Expand the definition of "sensitive covered data" in the American Privacy Rights Act (APRA) of 2024 to include brain data.[12]

**Rationale**

The US lags behind its international counterparts in sensitive data protection and healthcare cybersecurity. For example, the US lacks sweeping federal data protection legislation comparable to the GDPR of the European Union (EU), which has well-codified provisions and stricter cybersecurity regulations.[68] The US Senate's proposed APRA discussion draft serves as the first bipartisan federal data privacy bill and aims to mirror the GDPR.[69]

At the state level, specifically concerning neural data, Colorado has led the charge, passing HB 24-1058 in April 2024, which expanded the scope of the Colorado Privacy Act to protect neural data.

[12] At the time of writing, APRA is a proposed federal data privacy law. It was introduced by Congress in April 2024 and is currently undergoing the legislative process. It has not yet been enacted into law.

This makes it the first US state to cover sensitive brain data under existing consumer privacy law.[70] The APRA discussion draft is thus timely. APRA is envisioned to function independently of, yet complementarily to, state-level consumer privacy laws.[71] Expanding the definition of "sensitive data" to mirror the recent amendments introduced by the Colorado Privacy Act thus satisfies the goals of APRA. Protecting sensitive mental health data under the APRA aligns with the principles of non-maleficence and autonomy by preserving individual privacy rights and adhering to robust international standards in health data protection.

### 6.5. Brain hacking

*Brain hacking* is defined as the "emerging possibility of malicious actors accessing BCIs and other neural devices to compromise the operations of these devices."[5] Brain hacking requires breaching what Denning, Matsuoka, and Kohno (2009) call "neurosecurity"—a concern they identify as critical to the development and deployment of neural implants outside of contained research environments.[61] Neurosecurity is not sufficiently subsumed under existing cybersecurity regulations because the risks of "neuro-cyber" attacks differ from those posed by traditional cyber-attacks.[61]

**Recommendation**

(7) Amend Section 3305 of the FT&C Act to expand Section 524B "Ensuring Cybersecurity of Devices," and include an explicit provision for "Neurosecurity."

**Rationale**

Neurosecurity surpasses traditional definitions of cybersecurity, extending beyond physical harm and manifesting as a threat to an individual's personhood or identity.[61] First, the potential harms arising from neuro-cyberattacks are primarily emotional. Malign actors could "hack" the brain and elicit unwanted physical actions on the part of the BCI user—the loci of these actions threatening our societal understanding of the brain as a private haven.[5] Second, these potential harms threaten one's legal personhood and the protections afforded to individuals as legal citizens of a state. In a court of law, this may directly infringe upon one's constitutional protections.[5] Including an explicit provision for "neurosecurity" under Section 524B of the FT&C Act addresses the principles of non-maleficence and autonomy by protecting individuals from potential cyber threats that could compromise their mental integrity.

### 6.6. BCI-mediated action in criminal law

Brain-controlled prostheses permit a wide variety of therapeutic benefits for functionally impaired patients, restoring motion to both amputated zones and paralyzed limbs.[72,73] Despite improving patient quality of life, new legal ramifications arise from human-computer neural interfacing.[5,74] In traditional criminal law, *actus reus* is "the act or omission that comprises the *physical* elements of a crime as required by statute."[75] BCI-mediated action does not satisfy the physical criterion of *actus reus* in a traditional legal sense. Instead, any movement is triggered by neural activity, excluding BCI-mediated action from legal culpability by creating a potential loophole where any crime committed with a brain-controlled prosthetic limb, for example, could be exempt from prosecution.[74]

**Recommendation**

(8) Change the legal definition of *actus reus* to refer to bodily movements *as originating from neural action* (focusing on the locus of the action rather than its physical outcome).

**Rationale**

Criminal liability in cases of BCI-mediated action is complicated by the legal reference to the "physical elements" of a crime and the insufficiency of "involuntary action"[75] to satisfy *actus reus* requirements. This "loophole" lends itself to the possibility of a defence narrative for which precedent does not exist. For example, defendants could plead that their actions were a direct result of undue or involuntary mental influence (e.g., by "brain hackers" or even state actors) to be exonerated.

Liv (2021) notes that "recognizing neural activity as an indicator of movements will satisfy the factual basis of the offence," which is required to pass the two-pronged (factual and legal) causal test for a consequential offence.[5] Some concerns suggest that setting neural activity as the action-initiating requirement opens the possibility for unconscious or unintended thought to satisfy *actus reus* without the subject's awareness.[5] However, these concerns are not new; they overlap with the question of free will, which highlights the role of criminal law as a practice.[5] Clarifying the legal definition of *actus reus* addresses the principles of justice and autonomy by ensuring that individuals using BCIs are held accountable for their actions while recognizing the unique nature of BCI-mediated actions.

### 6.7. Psychological continuity

The psychological continuity of BCI users is central to discussions of personal identity and autonomy. Both disease and treatment can alter an individual's personality or sense of self, raising profound ethical and legal questions about the nature of personhood and the limits of medical intervention.[19] Furthermore, the dichotomy between personal identity as defined by one's *psychological* continuity versus one's *bodily* continuity raises interesting questions in the field of artificial implantation, particularly concerning BCIs. Hansson (2005) identifies potential changes to psychological continuity as arising from (i) the introduction of a foreign body into the brain itself, and (ii) physical complications caused by the surgical implantation or removal of the chip (e.g., tissue damage, scarring, or tumour formation).[19] If we consider bodily continuity a sufficient condition for continued personal identity, these two avenues present potential actionable areas for introducing regulation which explicitly protects psychological continuity.

**Recommendation**

(9) Codify psychological continuity as an emerging "neuroright" for protection under humanitarian law.

**Rationale**

In October 2022, the United Nations Human Rights Council adopted a resolution on neurotechnology, requesting action to protect and promote human rights, including novel recommendations for mental privacy, mental integrity, cognitive liberty, and psychological continuity.[76,77] While integrating neurorights into existing human rights frameworks provides a foundation for addressing these issues, we argue that a more comprehensive approach is necessary. Even if existing rights frameworks are sufficient to protect neuro*rights,* it is unclear how these protections can be extended to fit emerging neuro*technologies*. If bodily continuity has any effect on the continuation of personal identity, and if, according to Hansson, the very presence of an implant alters one's ability to think freely, then it is not clear how existing rights frameworks—like the right to freedom of thought—are meant to mitigate this change in identity and protect cognitive liberty.

The pioneering efforts of Chile in extending explicit constitutional protection over "neurorights" serve as a valuable model for other nations.[78] The Chilean amendment aimed "to give personal brain data the same status as an organ, so that it cannot be bought or sold, trafficked or manipulated."[79] By recognizing mental privacy, free will, and non-discriminatory equal access to neurotechnologies as fundamental rights, Chile has set a precedent for addressing the unique risks posed by emerging neurotechnologies. Establishing neurorights as independent rights with explicit protections and provisions encourages a more flexible and responsive legal framework capable of addressing the rapidly evolving landscape of neurotechnology. This approach adheres to the principles of autonomy and beneficence by safeguarding an individual's sense of self and ensuring that BCI treatments do not unduly alter a person's core identity.

Table 3 summarizes the ethical implications identified by considering the applications of next-generation BCIs, along with the recommendations advanced in their consideration.

Table 3: Summary – Applications

| Considerations for BCI *applications* | Recommendation(s) | Principle(s) addressed |
|---|---|---|
| Restoration of function to "normal threshold." | 1. Maintain quantitative threshold standards for neurophysiological illnesses with sufficient reference data and permit neurostimulation up until "normal" thresholds. | Justice. |
| | 2. Offer DBS by BCI-mediated action for treatment-resistant neuropsychiatric disorders, though not as a first-pass intervention. | Beneficence; Non-maleficence. |
| | 3. Couple BCI treatment with CBT, DBT, EMDR, or other forms of clinically accepted therapy under a licensed practitioner. | Beneficence. |
| Enhancement of function past "normal threshold," (including for entertainment). | 4. Restrict BCI-mediated enhancement to special permissions. | Justice; Non-maleficence. |
| Decoding processes. | 5. Advance stakeholder-specific device labelling requirements that focus on decoding functions. | Explicability; Autonomy. |
| Mental privacy. | 6. Expand the definition of "sensitive covered data" in APRA 2024 to include brain data. | Non-maleficence; Autonomy. |
| Brain-hacking. | 7. Amend Section 3305 of the FT&C Act to expand Section 524B "Ensuring Cybersecurity of Devices," and include an explicit provision for "Neurosecurity." | Non-maleficence; Autonomy. |
| BCI-mediated action in criminal law. | 8. Change the legal definition of *actus reus* to refer to bodily movements *as originating from neural action.* | Justice; Autonomy. |
| Psychological continuity. | 9. Codify "psychological continuity" as an emerging "neuroright" for protection under humanitarian law. | Autonomy; Beneficence. |

These applications highlight the complex interplay between technological capabilities (as realized by next-generation chip design), ethical considerations, and legal frameworks. As BCIs evolve, regulatory approaches must adapt to address these novel risks while striking a balance between the potential benefits and risks of these technologies.

## 7. Conclusion

The next generation of networked BCIs offers unprecedented therapeutic potential. However, these increasingly sophisticated devices also introduce new ethical and legal risks. Historically, the tight power parameters under which these devices must operate have created a trade-off between task flexibility and operational capacity. However, modern chip architectures (e.g., SCALO) are enabling new applications to become possible at scale. Despite their similarities to IMDs, next-generation BCIs raise new questions at the intersection of bioethics and digital ethics that existing regulatory frameworks are ill-equipped to address. Our analysis has identified key areas of concern in both the design and application of BCIs, including privacy, security, transparency, and equity issues.

In terms of design, we emphasize the need for a holistic approach that considers not only the technical aspects of next-generation BCIs but also their long-term implications for users and society. Regarding the application of BCIs, our analysis highlights the complex interplay between technological capabilities and fundamental human rights.

Our recommendations aim to provide a foundation for the development of comprehensive regulatory frameworks that can keep pace with the rapid technological advancements observed in BCI engineering. We emphasize the need for a multidisciplinary approach, bringing together experts from neuroscience, engineering, ethics, law, and policy to address these risks effectively. As BCIs continue to evolve and potentially expand beyond therapeutic applications, it is crucial that we proactively establish ethical guidelines and regulatory measures to ensure their responsible development and use. A proactive governance approach will protect individual patient rights and societal interests, while fostering public trust in these transformative technologies.

While next-generation BCIs hold immense promise for improving human health and well-being, their unique potential demands that we reevaluate our existing ethical and legal frameworks. By addressing these risks proactively, we can continue to deploy these technologies as life-sustaining or even life-saving treatments while safeguarding fundamental human rights, centring on individual values, and protecting patient interests.

*Appendix I*

Table 2: Comparison of "A Unified Framework of Five Principles for AI in Society" (Floridi and Cowls, 2019) and the principles of bioethics relevant to the ethical regulation of medical devices, as identified by the AMA.

| Principle | American Medical Association (AMA)[13] | Floridi and Cowls (2019)[9] |
| --- | --- | --- |
| Beneficence | "Aiming to do good for patients is the underlying motivation in solving any unmet clinical need." | "Promoting well-being, preserving dignity, sustainability." |
| Non-maleficence | "'First do no harm.' Most devices carry inherent risk, and the potential benefit must justify the potential risk." | "Privacy, security, 'capability caution.'" |
| Autonomy | "Respecting others' rights to make their own, fully informed choices demands that innovators be completely transparent with anyone affected by the technology, informing them of potential risks, benefits, and alternatives. It also demands disclosing all conflicts of interest." | "The power to 'decide to decide.'" |
| Justice | "Justice requires commitment to deciding fairly among competing interests, sometimes through third-party arbitration, in resolving conflict. It also calls for reasonable, nonexploitative, and well-considered procedures to be administered fairly." | "Promoting prosperity, preserving solidarity, avoiding unfairness." |
| Explicability | Not defined. | "Enabling the other principles through intelligibility and accountability." |

*Appendix II*

| Area of Limitation | Constraint |
|---|---|
| Chip Architecture | Traditional, monolithic ASIC design, with one processor per task, allows the performance of only one function (e.g., reversal of seizure onset).[7] |
| Power Expenditure | FDA, FCC, and IEEE safety guidelines limit available power to 15-40 mW.[4] |
| Flexibility | Generalizability requires high power, surpassing safe thresholds for heat dissipation. Low-power devices are inflexible, only supporting specific tasks in specific regions. |
| Therapeutic Potential | Broad adoption requires BCIs that can treat multiple neurological disorders (which may affect multiple brain regions). Multi-site BCIs need flexible processing algorithms under safe power limits.[4] |